%% file: main.tex
\documentclass[12 pt]{revtex4-1}

\usepackage[english]{babel}

\usepackage{amsmath}
\usepackage{amssymb}
\usepackage{graphicx}
\usepackage{braket}
\usepackage[colorlinks=true, allcolors=blue]{hyperref}

\begin{document}

\title{Data-driven Hamiltonian correction for qubits for design of gates}
\author{John George Francis$^1$ and Anil Shaji$^{1,2}$}

\address{$^1$ School of Physics, Indian Institute of Science Education and Research Thiruvananthapuram, Kerala, India 695551}

\address{$^2$ Center for High Performance Computing, Indian Institute of Science Education and Research Thiruvananthapuram, Kerala, India 695551}


\begin{abstract}
We obtain  correction terms for the standard Hamiltonian of 2 transmons driven by microwaves in cross resonance. Data is obtained from a real transmon system, namely ibm\_kyiv on the IBM quantum platform. Various data points obtained correspond to different microwave amplitudes and evolution times. We have an ansatz for the correction term and a correction operator whose matrix elements are parameters to be optimized for. We use adjoint sensitivity and gradient descent to obtain these parameters. We see a good fit in the predictions from the corrected Hamiltonian and hardware results demonstrating the effectiveness of scientific machine learning for fine tuning theoretical models to faithfully reproduce observed data on time evolution multiple qubit systems. 
\end{abstract}

\maketitle

\section{Introduction}

In its most abstract form, quantum information processing involves the application of an arbitrary unitary transformation on a register of qubits that holds the input data in order to generate the desired output~\cite{Feynman1982} \cite{NielsenChuang}. In practice, however, the unitary transformations that can be applied are limited to ones that can be broken up efficiently into a sequence of applications of elementary gates chosen from a universal set~\cite{dawson2005solovaykitaevalgorithm}. The circuit model of quantum computation~\cite{NielsenChuang} captures this idea and provides a pictorial language for constructing, visualizing and optimizing the arrangement of elementary gates for implementing a unitary transformation corresponding to a particular algorithm. In any physical realization of qubits and operations on them, it suffices to learn how to perform the universal gates accurately and efficiently since more complex information processing tasks can be performed by combining these operations according to the appropriate quantum circuit. 

Qubits can be formed either by learning how to control naturally occurring systems like ions, atoms, electrons, photons etc.~or by fabricating suitable devices like superconducting qubits, quantum-dot based qubits etc. These two broad approaches have their respective advantages and disadvantages \cite{Quantum_Computers_nature}. In this paper we are specifically interested in the mathematical description of the dynamics of the qubits furnished by their respective Hamiltonians. For naturally occurring qubits like atoms and photons, the respective Hamiltonians are typically simple and precisely known in most cases. For more complex systems like superconducting qubits and quantum dots, the corresponding Hamiltonians are also relatively complicated and not always precisely know. In addition, no two devices are completely identical and similar variability is ideally there in the Hamiltonians also~\cite{josephsonjn_reproducibility}. Even in the case of NV-center based qubits which are closer to the naturally occurring qubits than the devices, such variability is present due to the non-regular positioning and non identical neighborhoods of the defect centers~\cite{nv_quantum_computer}.

Knowing the Hamiltonian precisely allows one to design ideal external control or pulse sequences that implement the required gates with greater accuracy reducing the probability of gate errors. It also allows one to design control sequences specifically tailored for each quantum processor or device improving performance. A third advantage would be the ability to design control sequences that directly and natively implement more complex gates like multi-control rotations which, if available can reduce the circuit depth when running algorithms~\cite{oneshottoffoli}. An alternate approach would be to fine tune the control sequences empirically without assuming in-depth knowledge of the dynamics. On the other hand, with the knowledge of the Hamiltonian there are powerful methods based on AI/ML which can optimize complex control sequences rapidly and efficiently. In this paper we address the question whether AI/ML methods can be extended to learning either system-specific or device-specific corrections to the Hamiltonian of a few qubit system starting from a simplified version of the Hamiltonian which has previously been found to be effective and computationally tractable for design of control pulses. We prototype this approach on small scale quantum processors available through the cloud on IBM quantum \cite{IBMquantum}.

In our work we have utilized pulse-level control \cite{qiskit_pulse} of IBM quantum processors. Similar devices from other companies also allow similar level of control on their transmon qubits\cite{rigetti,OQC,IQM,QuantumCircuitsInc} which can be controlled and manipulated using microwave pulses. Offering this level of control over evolution of the state of the qubits unlocks the potential to optimize gates and implement gates outside the minimal universal set in one shot as mentioned earlier. Some earlier work on the design of the pulse sequences to implement various gates include a reinforcement learning approach~\cite{rl_control,rl_entangler}. Other, more traditional methods include GRAPE (Gradient ascent pulse engineering), CRAB (chopped random basis optimization), GOAT (Gradient optimization of analytic controls) and the Krotov method. A comparative study of some of these approaches can be found in \cite{comp_study}. 

In this paper we focus on learning the corrections to the Hamiltonian that governs the primary means of coupling and entangling two transmon qubits, namely the cross resonance gate. In experiments of simulating cross resonance we come across a mismatch between simulation and experiment which we quantify. This mismatch, as explained in \cite{effective_cr}, is predominantly due to cross talk. The outline of this paper is as follows. Section \ref{transmon} recaps the details of the quantum hardware, some previous work on theoretically modeling the qubits and other works which make use of the said hardware. Section \ref{the_problem} explains the problem we run into while trying to design gates using the theoretical model Hamiltonians that are typically used. Section \ref{hamil_correction} explains the method we use to use information about this mismatch to learn corrections to the Hamiltonian that was initially assumed. We gather a dataset and train our model to fit on a subset of the data. Section \ref{results} discusses the main results and the form of the correction matrix. A brief discussion of the main result and our conclusions are also included in this section. 

\input{transmon.tex}
\input{theproblem.tex}
\input{hamiltonian_correction.tex}
\input{results.tex}

\section*{Acknowledgments}
The authors acknowledge the support of the Center for High Performance computing at IISER TVM for use of the {\em Padmanabha} computational cluster. A.~S.~was supported in part by a grant from the Department of Science and Technology, Government of India, as part of the National Quantum Mission. 

\section*{Bibliography}
\bibliographystyle{iopart-num}
\bibliography{main}

\appendix
\input{appendix_plots.tex}

\end{document}

%% file: transmon.tex
\section{The Transmon Qubit}\label{transmon}

The IBM quantum platform provides pulse level control \cite{qiskit_pulse} of superconducting qubits which have the {\em transmon} architecture \cite{transmon}. We are specifically interested in a pair of such qubits that can be coupled directly to each other. Typical quantum processors available over the IBM quantum platform have many more qubits with direct interconnects between the qubits following specific topologies. For instance, the  IBM {\em Eagle} processor has 127 such superconducting qubits but we focus here on only a pair of such qubits. 

Transmon qubits are essentially Cooper-pair box type charge qubits in which two of the charge states of the system corresponding to different numbers of Cooper-pairs inside a superconducting island is taken as the qubit levels. The transmon architecture in which a basic Cooper-pair box is capacitively shunted allows for long coherence times. Charge states of the system form effectively an anharmonic oscillator and the anharmonicity allows for selection of two, out of the many levels, to form the qubit. A pair of such qubits system can be modeled as coupled Duffing Oscillators with independent external drives \cite{qiskit_dynamics} \cite{effective_cr}. The Hamiltonian for two adjacent qubits that are coupled can be written down as, 
\begin{eqnarray}
    \label{eq:H1}
    H & = & \sum_{i=1}^{2} \omega_i \hat{N}_i + \frac{\delta_i}{2}(\hat{N}(\hat{N}-I) + j_{12}(a_1^{\dagger}a_2 + a_1a_2^{\dagger}) \nonumber \\
    && \qquad + \Omega_1 S_1(t) \cos(\omega_{d1} t)(a_1^{\dagger}+a_1) + \Omega_2 S_2(t) \cos(\omega_{d2} t)(a_2^{\dagger}+a_2)
\end{eqnarray}
$a^{\dagger}$ and $a$ are ladder operators for each of the two qubits, $\omega_i$ is the natural frequency of qubit $i$, $\hat{N}$ is $a^{\dagger}a$, $j_{12}$ is the qubit-qubit coupling strength, and the drive strengths are given by $\Omega$. The factor $S(t)$ denotes a programmable time dependent scaling factor between -1 and 1 used to shape the pulse amplitude,  while $\omega_{di}$ are the frequencies of the microwave drives on each of the two qubits. We have assumed in the form of the Hamiltonian above that the microwave pulse couples only to the X-quadrature of each of the qubits. Transformation to a suitable interaction picture has also been employed in writing the Hamiltonian above to remove the modes of the microwave cavity that the transmons are placed in. In this sense, even at the starting point of our discussion, there are approximations that go into the Hamiltonian we write down and therefore under various scenarios corrections to the Hamiltonian are to be expected in addition.

The first three terms of the above Hamiltonian are time independent or static terms characterized by the parameters $\omega_i$, $\delta_i$ which are respectively the frequency and anharmonicity of the qubits and the coupling $j_{12}$ between the two qubits that. The coupling in the case of transmons depends, among other things, on the geometry of the microwave cavity housing the devices. The time dependent parts of the Hamiltonian pertains to microwave drives that can be applied independently on each of the qubits at respective frequencies $\omega_{d1}$ and $\omega_{d2}$. Manipulation of the qubits is achieved by controlling the drives.

The envelops, $\Omega_i S_i (t)$ of the drives can be programmatically modulated using the IBM qiskit pulse API. The phase of the pulses can also be controlled but we do not explore this aspect in the following. Single qubit rotations are carried out by applying drives of suitable amplitudes that are resonant with the  natural frequency of each qubit for specific durations. Entangling gates are implemented out by driving one the qubits in a pair of coupled ones at the natural frequency of the other \cite{crossresonance,effective_cr}. This {\em cross-resonance} (CR) gate is the only readily available, native two qubit coupling available in the transmon system. The qubit whose frequency is used for driving the other one is referred to as the {\em control} qubit. Under the cross resonance condition the drive term in the Hamiltonian is reduced to $\Omega(t) \cos(\omega_{2} t)(a_1^\dagger + a_1)$. To see how the CR Hamiltonian generates coupling between the two qubits, we first assume that only the two levels of the anharmonic oscillator that forms the qubit is considered even though other levels of the transmons can come into play depending on the strength and frequency of the drives and couplings. In other words we have considered the anharmonicity of the transmon levels to be effectively infinite. A slew of further approximations and simplifications leads to the following form for the CR Hamiltonian~\cite{effective_cr}:

\begin{equation}
	\label{eq:CRH1}
	H = \big(\Delta - \sqrt{\Delta^2 + \Omega^2} \big) \frac{Z {\mathbb I}}{2} - \Bigg( \frac{j_{12}\Omega}{\sqrt{\Delta^2 + \Omega^2}} \Bigg) \frac{Z X}{2}, 
\end{equation}
where $\Delta = \omega_1 - \omega_2$ and $\Omega$ is the drive strength in a frame co-rotating with the drive field with counter-rotating terms ignored under the rotating wave approximation. In the equation above, $X$ and $Z$ denote Pauli operators on the qubits. Eq.~(\ref{eq:CRH1}) is one of the more recent and widely used effective Hamiltonian models for the two transmon qubit system which was first described in terms of the microwave amplitude \cite{microwave_coupling} and \cite{crossresonance}. Detailed analysis of the two qubit system from various perspectives can be found in the literature. In  \cite{tuning_up_cr} it is shown how unwanted components in the CR hamiltonian can be minimized. This is the basis of the Echoed Cross Resonance gate. Equation (\ref{eq:H1}) is obtained in \cite{effective_cr} from a model of two transmons placed inside a resonator by integrating out the resonator. They further derive another effective hamiltonian which is block diagonal in terms of Pauli Operators and their coefficients (rates). Effects of cross talk between the qubits which introduces terms proportional to $IY$ in the effective hamiltonian are also discussed in \cite{tuning_up_cr}. In the context of cross talk between qubits it may be noted that the IBM backend does not report parameters for cross talk.

\subsection{Designing Pulse sequences for implementing gates}

If the Hamiltonian governing the evolution of a quantum system is known, one can utilise techniques from the well developed field of optimal control theory\cite{optimalcontrol} to design pulse sequences that implement desired gates on one or more qubits. Several frameworks like qiskit-dynamics \cite{qiskit_dynamics}, torchqc \cite{torchqc}  quantumcontrol.jl and qutip \cite{qutip} exist which allow simulating the time evolution of driven transmon qubits. These frameworks enable exploration of transmon dynamics without access to quantum hardware. Even if these tools allowing fast feedback and iteration, optimisation of pulse sequences for gates acting on multiple qubits as well as gates implementing complex transformations may still be computationally challenging. Use of techniques like GRAPE \cite{grape}, Krotov\cite{krotov}, CRAB\cite{crab} and machine learning help to rapidly converge on optimized pulse sequences for gates, state transfer and other protocols.

In  \cite{parametric_H} a method for reducing the computational challenges of designing optimal pulse sequences for transmon qubits starting from a model Hamiltonian is discussed. An interpolation method to reconstruct new pulses from an existing pulse sequence is introduced. The allows for implementation of new unitaries with time steps different from the existing pulse sequence that was used allowing for  pulse sequences that capture the dynamics under time varying Hamiltonians. The utility of this approach is demonstrated by simulating neutron scattering with using a quantum 4-level system. The underlying model of the transmon that is used in \cite{parametric_H} is the one mentioned in Eq.~\ref{eq:H1}.

Reinforcement learning is used in \cite{reinforcement_learning_CR} to design quantum pulses for entangling gates. The Reinforcement Learning algorithm is able to find a {\sc cnot} gate that is much faster than standard {\sc cnot} gates. While in principle, this approach is model free and can be used directly on any quantum hardware. However, in \cite{reinforcement_learning_CR} the method is demonstrated by learning against a quantum simulator and not using actual quantum hardware. The issue of substantial variation in gate fidelity for small variations in the Hamiltonian parameters is pointed out in this work. A more direct approach is taken in \cite{squeeze} which attempts to speed up existing gates by taking current pulse forms and shortening their duration while simultaneously tuning other parameters based on a live calibration of the qubits. It is found that naively squeezing the pulses increases the gate errors and an alternate way of squeezing pulses through pulse parameter particle filtering is implemented. 

As noted earlier, the pulse level optimisations that are done all require  accurate models of the transmons. Inaccuracies in the models typically mean that the results tend to be not so effective on real quantum hardware. In our efforts to design pulse sequences for simple two-qubit operations, we also find a discrepancies between the output expected from numerical simulations using Equation \ref{eq:H1} and the actual obtained obtained by running the pulses on IBM processors. Specifically we applied cross resonance pulses, i.e. driving qubit T (a target qubit) at the resonant frequency of qubit C ( a control qubit). This discrepancy, as highlighted in \cite{effective_cr}, may arise from cross talk as well as possible other coherent processes. Keeping specifically in mind the variability from device to device, we do not attempt to use more detailed or refined theoretical models but rather we device a machine learning based scheme for adding  correction terms to the hamiltonian in Equation \ref{eq:H1} to give an effective hamiltonian correction that is able to match the output from the quantum hardware much more accurately. 

%% file: theproblem.tex
\section{The problem statement}\label{the_problem}

\begin{figure}[!htb]
    \centering
    \includegraphics[scale=0.6]{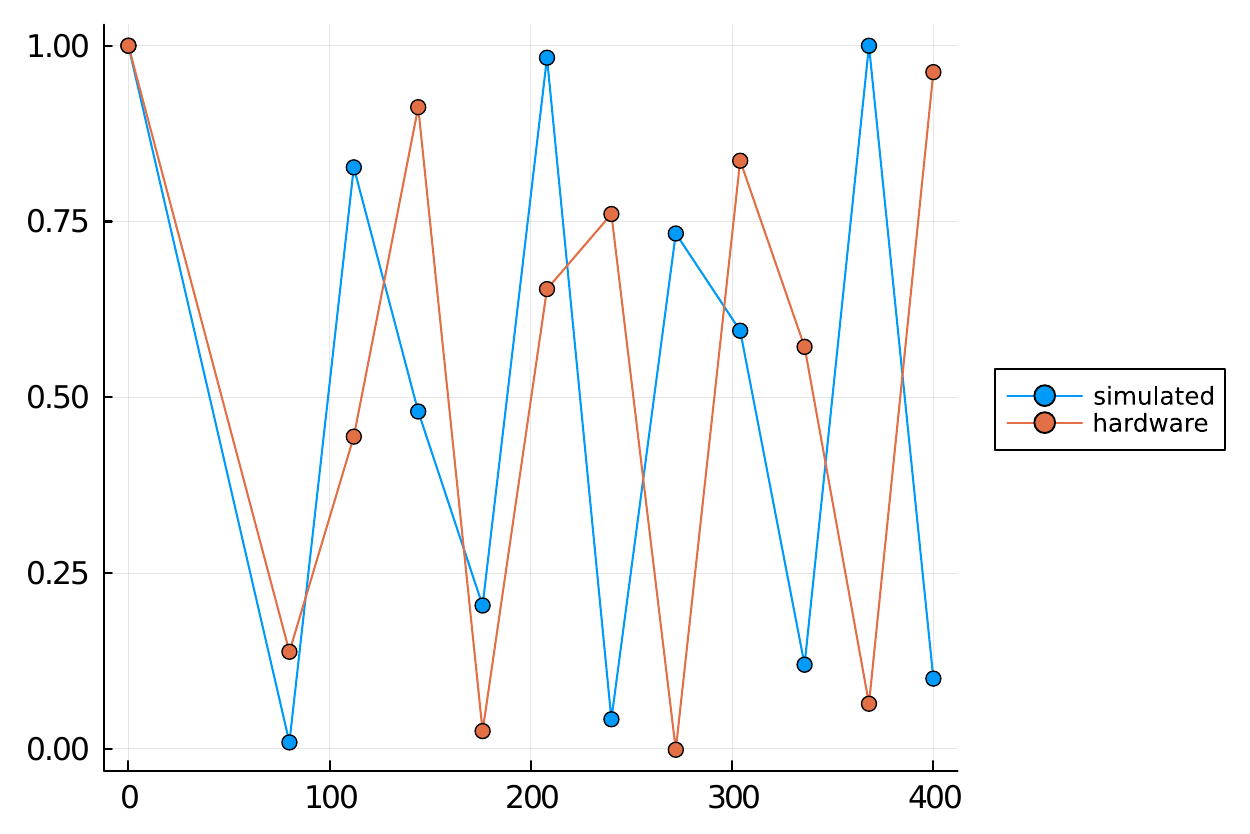}
    \caption{The survival probability of the initial state $\ket{000}$ at end of evolution using Gaussian square pules of different total duration as observed is shown by the red dots. The blue dots represent the same values computed by evolving the Schrodinger equation under the Hamiltonian given in Eq.~(\ref{eq:H1}). We see that the observed probabilities and the numerically computed ones diverge after very short times.}
    \label{fig:osaka_000}
\end{figure}

The problem is to model the coherent errors that arise from cross talk between qubits when  applying the cross resonance gate on a pair of transmon qubits in a three qubit system. Fig.~\ref{fig:osaka_000} shows the sequence of probabilities of finding the three qubits in the state $\ket{000}$ when specific microwave pulses on different durations are applied on the qubits. The plot shows both the expected probability obtained from simulating the evolution of the qubits using the Hamiltonian in Eq.~(\ref{eq:H1}) as well as the probabilities obtained from runs on quantum hardware. These runs were done on the \textit{ibm\_osaka} processor with both qubit 0 (counting starts from least significant bit and qubit 1 driven by a gaussian pulse at the frequency of qubit 1. in other words, qubit 1 is driven at its resonance frequency and qubit 0 is driven at cross resonance to 1. The third qubit, namely qubit 2 was a mere spectator and no pulse sequence was applied on it. The circuit was initialized with 3 qubits using qiskit. The data shown in Fig.~\ref{fig:osaka_000} corresponds to 11 different gaussian pulses all the same peak value but with increasing durations and standard deviations equal to one fourth of the duration. For each of the runs, the qubits were initialised in the state $\ket{000}$. The probability of finding the three qubits in the state $|000\rangle$ after the application of the pulses of various durations were measured on the \textit{ibm\_osaka} processor and it was also numerically computed using the model in Eq.~(\ref{eq:H1}). The values of parameters appearing in the Hamiltonian were extracted from the ibm qiskit API, which are updated regularly.

We see that there is substantial mismatch between the probabilities predicted by the simulation and the actual results. The actual runs may be affected by both coherent noise as well as incoherent noise due to external influences. Our focus is to find corrections to the Hamiltonian in Eq.~(\ref{eq:H1}) using the data collected from the actual runs and compensate for both types of errors to the maximum extent possible. Note that correcting the Hamiltonian is not typically expected to compensate for incoherent errors. However, the open quantum dynamics that leads to incoherent errors are typically described by GKSL type master equations whose generator may include a Hermitian part which can technically be compensated for by a Hamiltonian correction. 

%% file: hamiltonian_correction.tex
\section{Methodology}\label{hamil_correction}

Adjoint sensitivity is a technique to efficiently compute the derivatives of functionals of solutions of differential equations.  It has found several applications including  estimation of unknown parameters of Ordinary Differential Equations (ODEs) and finding more accurate models to describe observed data using Universal ODEs~\cite{rackauckas2020universal}. In reference ~\cite{rackauckas2020universal} it is shown that a discovery-improved model can be obtained from the data corresponding to a particular ODE (Lotka-Voltera model). The improved model is obtained by augmenting the known differential equation with a universal approximator which in the reference is a neural network. In general the approximator can be any construct that is parametrized and is capable of standing in for any possible function within the allowed ranges of the parameters.  In our case we explore augmenting the differential equation with up to three correction terms, one of them is static and remaining two are dynamic, time dependent terms. Since we are introducing the approximators as corrections to the Hamiltonian, they are also matrices of the same size. The independent elements of these matrices are the parameters that can be tuned to achieve better agreement with the observed data. 

The ansatz we use as the correction to the Hamiltonian is of the form 
\begin{equation}\label{H_c}
    H_c = \hat{H_1} + \hat{M} + S_1(t)cos(\omega_{1} t) \hat{D_1} + S_2(t)cos (\omega_{1} t) \hat{D_2},
\end{equation}
where $\hat{H_1}$ is the original Hamiltonian from Eq.~\ref{eq:H1}, $\hat{M}$ is a Hermitian $d^2 \times d^2$ matrix with $d$ being the number of levels of the transmon we use in the simulation. Similarly $\hat{D}_1$ and $\hat{D}_2$ are also Hermitian matrices of the same size representing dynamic corrections to the drive. In the present case use three levels for each transmon and eventually trace over the additional level which is outside the qubit space when computing the probabilities that form the output of the simulations. The solutions of interest describe the time evolution of the transmon qubits obtained by numerically integrating the Schrodinger equation using a suitable model and its Hamiltonian. The functional of interest is a cost function characterizing the deviation of the numerical solution from the observed evolution. We use the adjoint sensitivity \cite{SciMLBook, NODE, rackauckas2020universal} method to correct Hamiltonian by tuning all the independent elements of the three correction matrices. The task is to find some $M$ and $D_i$'s which explains the data obtained from running the pulses on the IBM device.

\subsection{Training dataset}

The first step in implementing the correction is creating a training dataset. A single datapoint has three elements. The first is the pulse sequence which determines the Hamiltonian apart from the correction terms, the second is a chosen initial state for the transmons and the third is the probabilities for various computational basis states of the transmons obtained from numerically evolving the system with this pulse  initial state. These probabilities are computed at the end of the pulse so that they can be compared with the probabilities obtained from actual device runs. 

Each pulse is a Gaussian square pulse on both qubits with amplitudes denoted by $A_1$ and $A_2$ respectively. $A_1$ is chosen from the set \{0.0, 0.01, 0.02, 0.03, 0.04\} (5 possible values) while $A_2$ is chosen from the set \{0.1, 0.2, 0.3, 0.4, 0.5, 0.6\} (6 possible values). The range of these values is selected similar to the values used in Qiskit's cross resonance gate. $A_2$ is the amplitude of the pulse on qubit 2 and drives the cross resonance, we want this to dominate over the on-resonant rabbi oscillation induced by $A_1$. For each these 30 possible amplitude pairs 20 pulses of increasing duration were generated. The pulse durations start from $320\,dt$ and increase in in increments of $128\,dt$, where $dt = 0.22222222$ nanoseconds is the resolution of the Arbitrary Waveform Generator (AWG) that is part of the {\ttfamily ibm\_kyiv} system. The pulses have an additional $32 \,dt$ gaussian rise and falls on both ends. Hence the durations range from $(320+64)\, dt = 85.33$ ns to  $(320+19\times128+64)\, dt = 625.77$ ns. An exception is the case of the amplitude pair ($A_1$,$A_2$) = (0.0,0.1) where we have 30 pulses, ranging from $(320+64)\, dt$ to $(320+29*128 + 64)\, dt$ in increments of $128\, dt$. This is because for this small amplitude, we find that a slightly larger evolution time was necessary for getting accurate results. We chose two different initial states for the two qubits on interest, $\ket{00}$ and $\ket{10}$ and for each initial state computed the probabilities for all possible final states ($\ket{00}$, $\ket{01}$, $\ket{10}$ and $\ket{11}$) for all the possible pulse shapes and amplitudes considered.

A preliminary test done using a few amplitudes using all four computational basis states as initial states indicated that learning on only two initial states, namely, $\ket{00}$ and $\ket{10}$ was sufficient to eventually predict correctly the time evolved probabilities corresponding to the remaining two possible starting states $\ket{01}$ and $\ket{11}$. In the interest of reducing our consumption of limited time available on IBM QPU's we only generated five data points with starting states $\ket{01}$ and $\ket{11}$. This was just to compare if the simulations with the corrected Hamiltonian would work on these initial states as well. The durations of these pulses were chosen to be the five largest durations in the pulses used with the other two initial states having durations starting from  $(320+15*128+64)\,dt$ to $(320+19*128+64)\,dt$ in increments of $128\,dt$.

We use part of the data available for every amplitude pair for training.  For a particular amplitude pair we have 20 possible pulse durations paired with corresponding output probabilities for the initial state $\ket{00}$. Similarly there are 20 data points with initial state $\ket{10}$, 5 with initial state $\ket{01}$ and 5 with $\ket{11}$. Out of these we use only the first 10 data points with initial states $\ket{00}$ and $\ket{10}$ for training. The remaining are used for validation and testing. The training set is then the target probability vectors labeled as $P^{00}_1, P^{00}_2 ... P^{00}_{10}$, $P^{10}_1, P^{10}_2, ... P^{10}_{10}$ along with the corresponding pulses and initial states that generated them. The superscript on the probability labels indicate the starting state while the subscript corresponds to increasing order of pulse duration starting from the shortest. We simplify the labeling of the 20 sets as $P_1...P_{20}$. Note that each of the $P_j$'s represents a probability vector for finding the state of the two transmons in any one of the four possible basis states. These probability vectors are obtained from actual device runs. Now for the same pulse parameters and initial states, we simulate the time evolution of the transmon qubits using the Hamiltonian in Eq.~(\ref{H_c}). From the final state computed in this manner, we  obtain the simulated probability vectors labeled as $Ps_1,Ps_2 ... Ps_{20}$.  Our objective is to find the Hamiltonian corrections such that the probability vectors obtained from the simulations match the observed probability vectors.

\subsection{The loss function}

The parameters we have to fix using the training dataset are the  $3d^2$ elements of the matrices $\hat{M}$,$\hat{D_1}$ and $\hat{D_2}$. Here $d$ is the dimension of the Hilbert space of the transmons we are considering. These parameters can be considered as elements of a flattened array  $\vec{p}$ of length $3*d^2$. The three Hermitian operators are constructed from $\vec{p}$ by taking $d^2$ of them at a time, arranging them into $d\times d$ matrices and taking the average of the matrix so formed and its adjoint to ensure that the resultant correction matrix is Hermitian. We start with all elements of $\vec{p}$ equal to zero which means that we are starting from the theoretically obtained Hamiltonian from Eq.~(\ref{eq:H1}).  We are interested only in corrections to the non-diagonal entries of the Hamiltonian since the diagonal entries are the fixed transmon energy levels. The diagonal elements of the corrections matrices are forced to be zero by always setting them to 0 during each step of the training process. As we will see later, this ensures that  gradients with respect to the parameters that determine the diagonal elements of the correction matrix as well as the diagonal elements  themselves are always 0. These parameters therefore do not change during the training process.

To quantify the performance of the correction applied we have to define a loss-function which will be minimised by the training process. The loss-function that we use has the form,
\begin{equation}
	\label{lossfunc}
		L(p) = \sum_i^{20} L_i(|\psi(T_i,p)\rangle) =  \sum_i^{20} |P_i - Ps_i(|\psi(T_i,p)\rangle)|
\end{equation}
In the equation above, $|P|$ denotes the sum of the absolute values of the elements of the vector $P$ while $|\psi(t,p)\rangle$ is the state of the transmon qubits obtained time evolving the respective initial states till time $t$ using a Hamiltonian that contains the correction parameters $p$.  $T_i$ is the total time duration of the $i^{\rm th}$ pulse. Minimizing $L$ to a value near 0 gives us the correction matrices $\hat{M}$, $\hat{D_1}$ and, $\hat{D_2}$ that, in principle, will fit the simulation to the training data points. As noted earlier, we find that training on 10 data points each with starting states $\ket{00}$ and $\ket{10}$ is sufficient to recover the corrections that reproduces all observed datapoints for a particular amplitude pair. This is then repeated for different amplitude pairs and we obtain the correction matrices for every amplitude pair in the dataset.

To minimize $L$ we use gradient descent which requires us to compute $\partial L/\partial \vec{p}$. In order to compute $L(p)$ we have to solve a differential equation (The Schrodinger equation) to obtain the state at the end of the pulse from which one can compute the probabilities for the system being in each of the computational basis states. In other words, $L(p)$ depends on the final value of the solution of an ODE initial value problem. An efficient means of obtaining $\partial L/\partial \vec{p}$ is by solving an adjoint differential equation. Specifically, we use the adjoint sensitivity approach to find $\partial L/\partial \vec{p}$ which is outlined below. 

\subsection{Adjoint sensitivity}

Consider an an initial value problem,  
\[ \dot{y} = f(y,p,t), \qquad  y(0) = y_0, \]
where $y$ and $p$ can, in general, be arrays of scalars rather than just a scalar function and a corresponding parameter. Let the solution of the initial value problem be $y(t,p)$ and let $g(y(t,p))$ be a real-valued function of the solution of the ODE at a particular time. Consider a functional $G$ of $g$ which is also real-valued and defined as
\[ G[y] = \int g(y(t,p),t) \, dt. \]
Using adjoint sensitivity one can compute the derivative $\partial G/\partial p$. For the case we are interested, rather than having a function $g(y)$ of the full solution $y(t,p)$ we are only interested in $g(y)$ computed using the solution $y(T,p)$ computed at one or more discrete time points denoted by $T$. Discrete adjoint sensitivity is used when the functional $G$ depends on the solution of the ODE at particular time points. For such problems $g(y(t,p),t)$ will be a convolution of the loss function with suitable delta functions. In our case the ODE is the time dependent Schrodinger equation and the functional we try to minimize over is  ,
\begin{equation}
\label{eq_L}
L_i[|\psi(T_i,p)\rangle] = \int_{t_0}^{T_i} L_i(|\psi(t,p)\rangle) \delta(t-T_i) dt
=  \int_{t_0}^{T_i} |P_i - Ps_i(|\psi(t,p)\rangle)| \, \delta(t-T_i) dt
\end{equation}
The derivative, $\partial L/\partial p$ is then obtained as
\begin{equation}
	\label{Lder}
    \frac{\partial L}{\partial p} =  \lambda^* (t_0) \frac{d|\psi(t,p)\rangle}{dt}\Bigg|_{t_0} + \int_{t_0}^{T_i} \left(\frac{\partial g}{\partial p} + \lambda^* \frac{\partial f}{\partial p} \right)dt,
\end{equation}
where $g(|\psi(t,p),t) =|P_i - Ps_i(|\psi(t,p)\rangle)| \, \delta(t-T_i)$ and $f=-i\frac{H(p)}{\hbar} \ket{\psi(t)}$ with $H(p)$ being the Hamiltonian with correction terms determined by $p$. In Eq.~(\ref{Lder}),  $\lambda$ is the solution to the ODE problem 
\[    \dot{\lambda} =  - \frac{df}{dy}^* \lambda + \frac{dg}{dy}^*  \quad {\rm with} \quad  \lambda(T) = 0,  \]
 with $y=|\psi(t,p)\rangle$. We obtain $\partial L/\partial p$ is then summing up the gradients $\partial L_i/\partial p$. These gradients can then be used in any gradient based optimizer to minimize $L$. A comprehensive description of the adjoint sensitivity method can be found in ~\cite{SciMLBook}.

\subsection {Tools and packages}

We implemented the integration of the Schrodinger equation for the transmons using the Hamiltonian including the correction term in the Julia programming language. The simulation incorporates the initial states, the various operators that are used in the Hamiltonian in Eq.~(\ref{eq:H1}), the pulses and parameters determining the energies, couplings, correction terms etc. The system parameters like the couplings can be directly imported from the IBM backend. Time evolution of the system is done using DifferentialEquations.jl \cite{DifferentialEquations.jl}. We use SciMLSensitivity.jl \cite{rackauckas2020universal,sense_alg} Julia library for computing the adjoint sensitivities. Optimisers.jl is a Julia package that implements several gradient based optimisers. It is a part of Flux.jl \cite{fluxjl}, a machine learning package in Julia. We use its implementation of the Nesterov \cite{nesterov} momentum optimizer. An input to the adjoint sensitivity is the set of derivatives of the loss function with respect to the ODE state $|\psi(t,p\rangle$. It is possible to obtain analytic expressions for these derivatives but our purpose is better served by automatic differentiation. We use Enzyme.jl a Julia extension to the Enzyme autodiff library\cite{enzymejl} for its speed and flexibility of differentiating arbitrary Julia code. The minimization leading to the correction matrices for the Hamiltonian is quite fast and takes around 15-20 minutes on a reasonably recent personal computer (The optimization was done on laptop computer AMD Ryzen 5 5600U with 16GB of RAM).

%% file: results.tex
\section{Results and Conclusion}\label{results}

The optimization procedure described in the previous section lead to the matrices $\hat{M}$, $\hat{D_1}$ and $\hat{D_2}$ that are corrections to the Hamiltonian in Eq.~(\ref{eq:H1}). We find that leaving out $\hat{D_1}$ and $\hat{D_2}$ from the simulation will lead to a static correction that is able to explain the evolution. Alternatively, leaving out $\hat{M}$ results in the dynamic corrections $\hat{D_1}$ and $\hat{D_2}$ also explaining the evolution faithfully. A third alternative was to have only $\hat{D_2}$ as the correction and in this case also fairly good fit to the observed data is obtained. The most general form of the correction that we took which included both static and dynamic corrections appears to be more than what is required to obtain agreement with experimental runs. It suffices to consider fewer correction terms and typically corrections to the dynamic parts are easier to implement since they are related to modifying the applied pulse sequence. We discuss below in detail the third case which is the minimal one with only one correction term to the drive on the second qubit of the form $cos(\omega_2 t) \hat{D_2}$. 

\begin{figure}[!htb]
   \centering
   \includegraphics[scale=0.6]{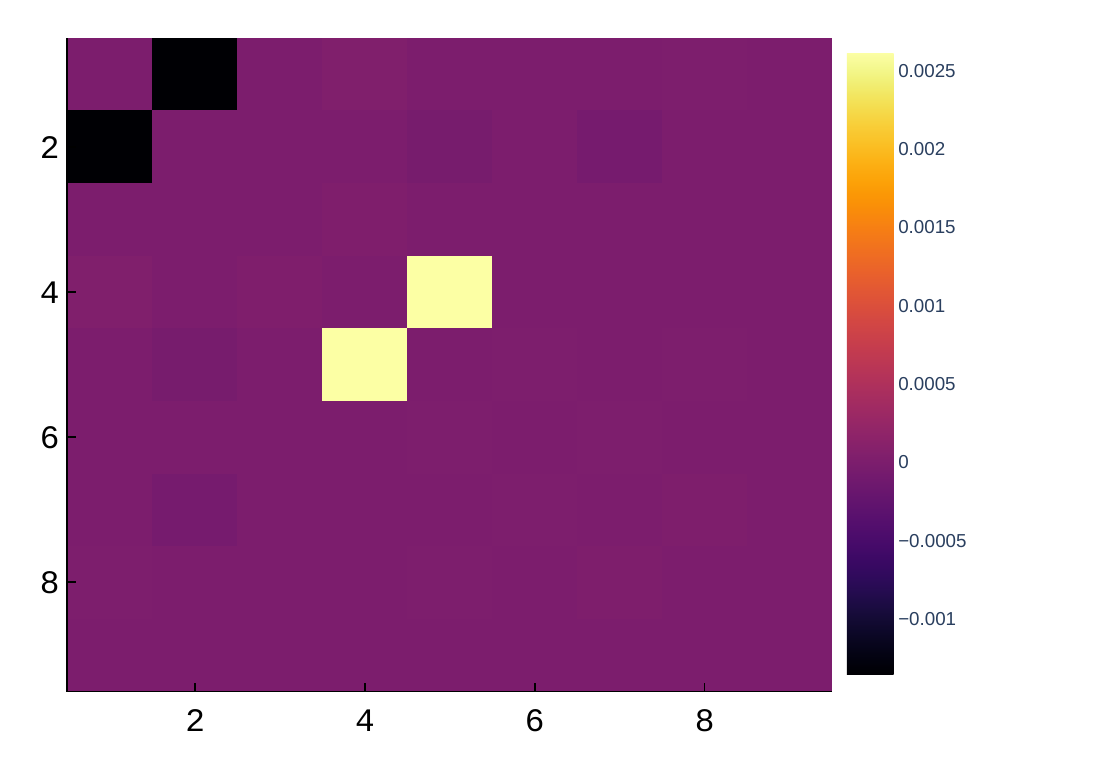} 
   \caption{heatmap of the correction matrix $\hat{D_2}$, elements (1,2), (2,1) (4,5) and (5,4) corresponding to the transitions $\ket{00}\bra{01}$ and $\ket{10}\bra{11}$ are non zero, while other elements are close to 0. Note that we are considering 3 levels for each transmon hence $\hat{D_2}$ is a $3^2\times3^2$ matrix.}
   \label{fig:heatmap}
\end{figure}

We considered three levels of the transmon when numerically integrating the Schrodinger equation. For a pair of transmons, the Hamiltonian and the correction were therefore $9 \times 9$ matrices for our case.  The heat map of a typical correction matrix that we obtained is shown in Fig.~\ref{fig:heatmap}. While most of the elements are close to 0, the matrix elements corresponding to the transitions $\ket{00}\bra{01}$ and $\ket{10}\bra{11}$ and their hermitian conjugates are significant. 

\begin{figure}[!htb]
    \centering
    \resizebox{7.5 cm}{5.2 cm}{\includegraphics{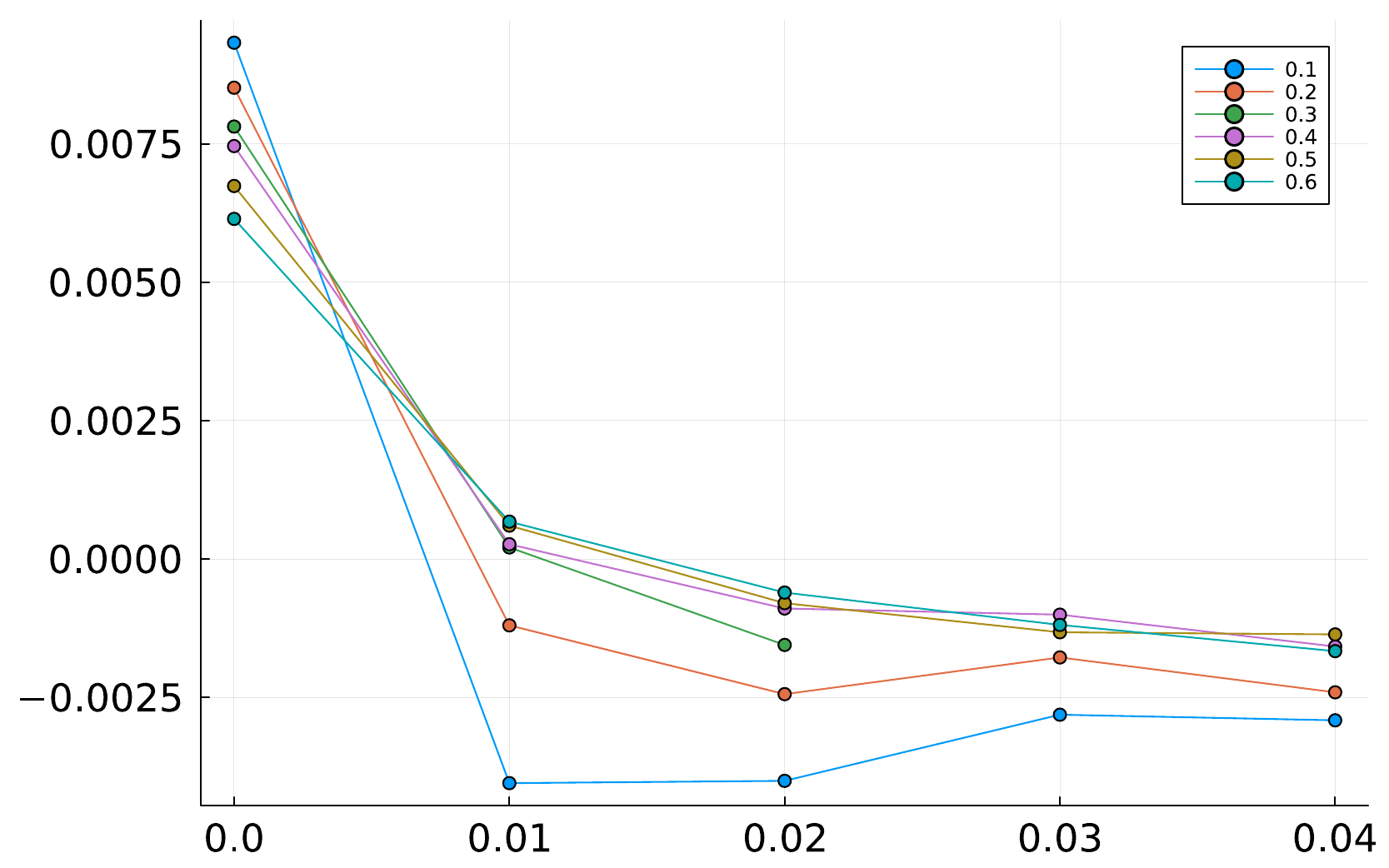}}
    \resizebox{7.5 cm}{5.2 cm}{\includegraphics{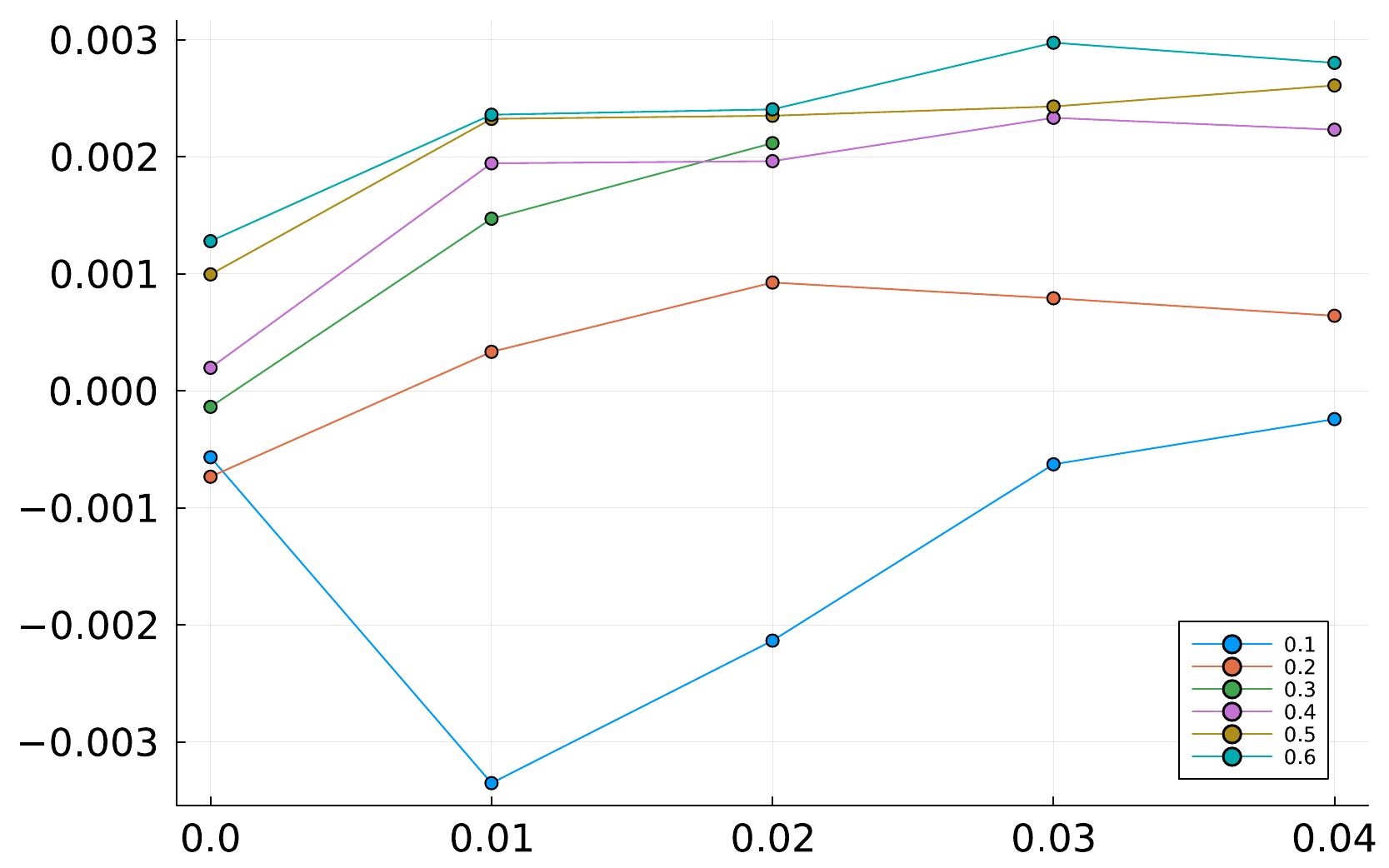}}
    \caption{The plot on the left shows the variation of $\bra{00}\hat{D}_2\ket{01}$ and the plot on the right shows the variation of $\bra{10}\hat{D}_2\ket{11}$ with respect to the amplitude of the applied pulse. Different lines correspond to the cross resonance amplitude on the control qubit while the $x$ axis shows the amplitude on the target qubit}
    \label{fig:mat12_plot}
\end{figure}

In Fig.~\ref{fig:mat12_plot} we plot the variation of the two significant matrix elements of $\hat{D}_2$ namely $\bra{00}\hat{D}_2\ket{01}$ and $\bra{10}\hat{D}_2\ket{11}$ as a function of the amplitude of the pulses applied on the control and target qubits. We see that while $\bra{00}\hat{D}_2\ket{01}$ decreases with the amplitude, $\bra{10}\hat{D}_2\ket{11}$ shows an increasing trend, indicating that there is an optimal choice of amplitude for which the correction required is minimal. The amplitudes used in empirically optimised pulse sequences for implementation of various gates on the quantum processors fall in this range. 

\begin{figure}[!htb]
    \centering
    \resizebox{7.5 cm}{5.2 cm}{\includegraphics{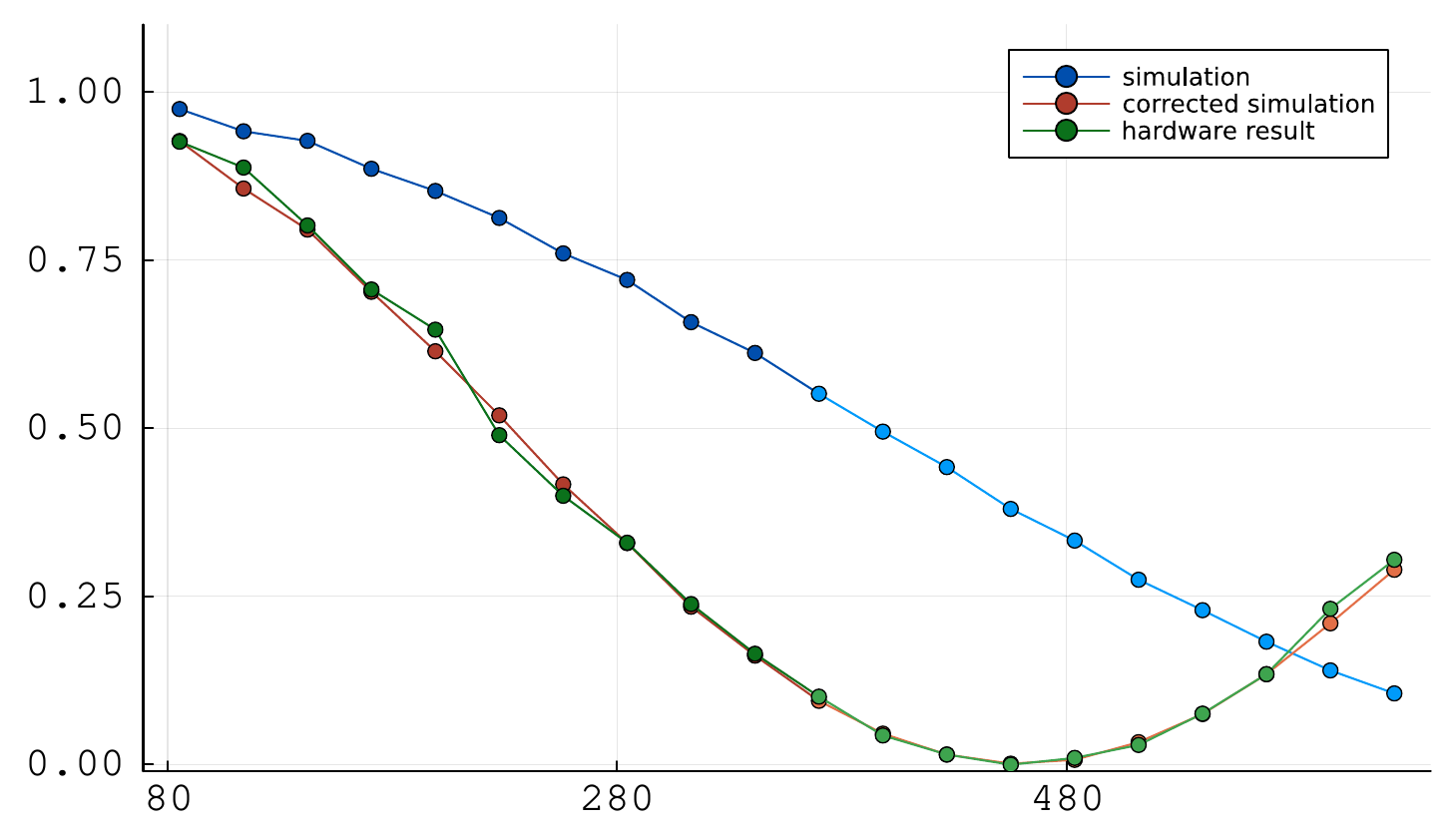}}
    \resizebox{7.5 cm}{5.2 cm}{\includegraphics{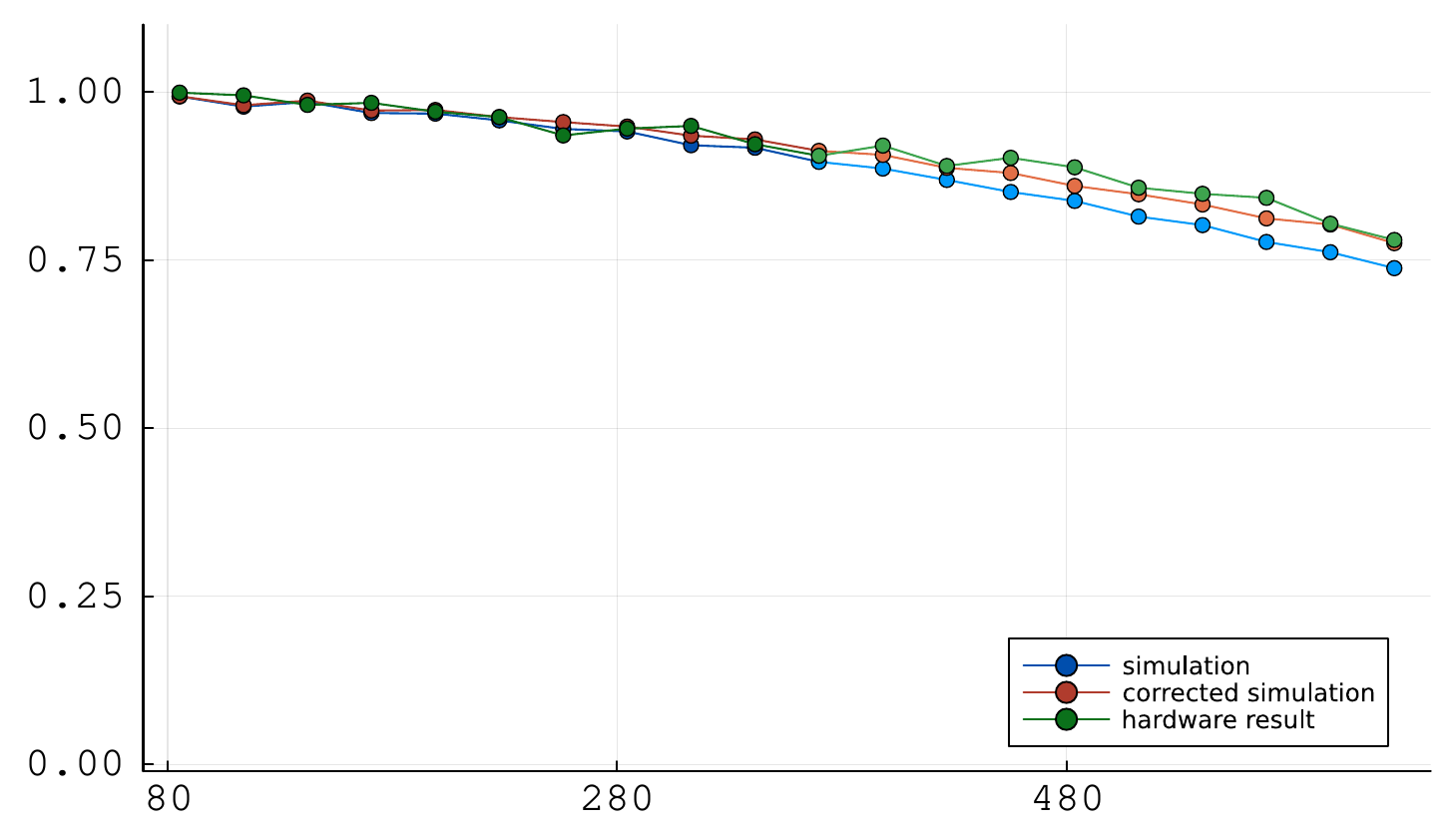}}
    \caption{The figure on the left shows the time evolution of probability $|\bra{00}\ket{\psi}|^2$ with the qubits initialized in the state $\ket{00}$   while the figure on the right shows $|\bra{10}\ket{\psi}|^2$ initialized in the state $\ket{10}$. The $x$-axis shows time in nanoseconds. The green dots show the data from the device runs while the blue dots show the probabilities obtained from numerical integration of the Schrodinger equation using the uncorrected Hamiltonian. The orange dots show the probabilities obtained using the corrected Hamiltonian showing very good agreement with the device run data. Darker shaded points correspond to points from the training set. Lighter shaded points were not used in training. The amplitude is $0$ on the target and $0.4$ on the control for both the plots.}
    \label{fig:evolution}
\end{figure}

In order to examine the effectiveness of the corrections obtained we compare the time evolution of different initial states computed using the corrected and uncorrected Hamiltonians with the data from the device runs. Note that only half of the device run data was used for training but for testing and comparison, all 20 time stamps were used for each of the initial states $\ket{00}$ and $\ket{10}$ as well as for each of the amplitude combinations. The last 10 points in each of the graphs plotted in  Fig.~\ref{fig:evolution} correspond to data that was not used for training. We see that for both initial states, the corrected Hamiltonian performs significantly better than the uncorrected one with very good agreement between the device run data. In the case where the initial state is $|10\rangle$, we see from Fig.~\ref{fig:evolution} that the agreement between data and the numerical evolution using the uncorrected Hamiltonian is also good with the correction improving matters only slightly. However, for this initial state also when the amplitudes of the pulses are increased the disparity between the data and uncorrected evolution increases. Corresponding plots for all the amplitudes are included in \ref{appendixA}.

The loss function given in Eq.~(\ref{lossfunc}) relative to the data from the device run was computed for the numerically obtained probabilities using both the uncorrected and corrected Hamiltonians. This was done for various choices of amplitudes of the drives for both choices of initial states. The losses are tabulated in Table \ref{tab:losstable}. With the uncorrected Hamiltonian we see a wide range of average error in computed probabilities with values ranging to as high as 0.48. On the other hand when the corrected Hamiltonian is used, irrespective of the drive amplitudes we obtain values of the loss function which are less than 0.05. 

We also carried out optimizations on a 3 qubit case where 2 qubits are driven at the frequency of the 3rd for a single instance of amplitude triplet. This is done with the aim of optimizing the pulse sequences after obtaining the required corrections for implementing three qubit gates like the Toffoli gate as native gates using pulse level control while improving their efficiency.  We find promising improvements in preliminary investigations but the losses were higher than in the 2 qubit case. Lack of sufficient data from device runs was the main impediment in improving these results. Investigations in this direction are left as future work in view of the time taken to obtain device runs on the hardware available over the cloud. 

\begin{table}[!ht]
    \centering
    \begin{tabular}{|c|c|c|c|}
    \hline
    \ \ \  T$_{\rm amp}$ \ \ \  & \ \ \  C$_{\rm amp} $ \ \ \  & Loss (uncorrected) & Loss (Corrected)\\
    \hline
    0.0 & 0.1 & 0.285 &  0.024 \\ 
    0.0 & 0.2 & 0.265 &  0.020 \\ 
    0.0 & 0.3 & 0.330&   0.029 \\ 
    0.0 & 0.4 & 0.301&   0.028 \\ 
    0.0 & 0.5 & 0.373&   0.029 \\ 
    0.0 & 0.6 & 0.483 &  0.030 \\ 
    \hline
    0.01 & 0.1 & 0.069 &  0.033 \\ 
    0.01 & 0.2 & 0.047 &  0.027 \\ 
    0.01 & 0.3 & 0.075 &  0.028 \\ 
    0.01 & 0.4 & 0.130 &  0.036 \\ 
    0.01 & 0.5 & 0.186 &  0.035 \\ 
    0.01 & 0.6 & 0.228 &  0.038 \\ 
    \hline
    0.02 & 0.1 & 0.071 &  0.028 \\ 
    0.02 & 0.2 & 0.071 &  0.031 \\ 
    0.02 & 0.3 & 0.101 &  0.042 \\ 
    0.02 & 0.4 & 0.123 &  0.029 \\ 
    0.02 & 0.5 & 0.168 &  0.035 \\ 
    0.02 & 0.6 & 0.194 &  0.038 \\ 
    \hline
    0.03 & 0.1 & 0.059 &  0.039 \\ 
    0.03 & 0.2 & 0.076 &  0.038 \\ 
    0.03 & 0.4 & 0.146 &  0.031 \\ 
    0.03 & 0.5 & 0.182 &  0.042 \\ 
    0.03 & 0.6 & 0.234 &  0.034 \\ 
    \hline
    0.04 & 0.1 & 0.056 &  0.036 \\ 
    0.04 & 0.2 & 0.082 &  0.048 \\ 
    0.04 & 0.4 & 0.156 &  0.036 \\ 
    0.04 & 0.5 & 0.209 &  0.031 \\ 
    0.04 & 0.6 & 0.258 &  0.041 \\ 
    \hline

    \end{tabular}
    \caption{Loss function values relative to the data from the device runs obtained for numerically computed probabilities using the corrected and uncorrected Hamiltonians.  T$_{\rm amp}$  denotes the amplitude of the Gaussian Square pulse applied on the target qubit while  C$_{\rm amp}$ is amplitude on the control qubit. }
    \label{tab:losstable}
\end{table}

In conclusion we find that applying Machine Learning techniques provide an efficient and quick way in which to augment the theoretical models that allow for efficient optimization of pulse sequences for implementing specific gates. The corrections found in this manner do not increase the complexity of the pulse optimization problem. This is unlike the cases where direct refinement of the model is done through the addition of analytic expressions in the Hamiltonian for higher order effects because in such cases the complexity of numerical integration of the Schrodinger equation is also typically increased. The efficiency of our machine learning algorithm opens the possibility of learning the relevant corrections in real time for each quantum processing unit and dynamically optimizing the pulses to obtain better performance. This can be of use in a scenario where quantum processors are mass produced but with slight individual variations like in the case of defect center qubits for which pulse sequences for elementary gates, tailor made and optimized for each qubit in each unit can be designed and implemented to ensure uniform performance despite the variations.

%% file: appendix_plots.tex
\newpage 

\section{}\label{appendixA}

 The plots comparing the observed data with evolution under the uncorrected Hamiltonian in Eq \ref{eq:H1} and corrected Hamiltonian for different amplitudes of the Gaussian square pulses are given below for the two initial states we consider. in all the four plots given below, blue lines show the survival probability obtained using the uncorrected Hamiltonian, green lines are data points obtained from device runs and orange shows the probabilities obtained from the corrected Hamiltonian. The $x$ axis is time in nanoseconds and the $y$ axis is population fraction. In the plots  $T$ denotes the target qubit pulse amplitude and $C$ is the control qubit pulse  amplitude in arbitrary units.
 
\begin{figure}[!htb]\label{fig:allplots00}
    \centering
    \includegraphics[width=0.99\linewidth]{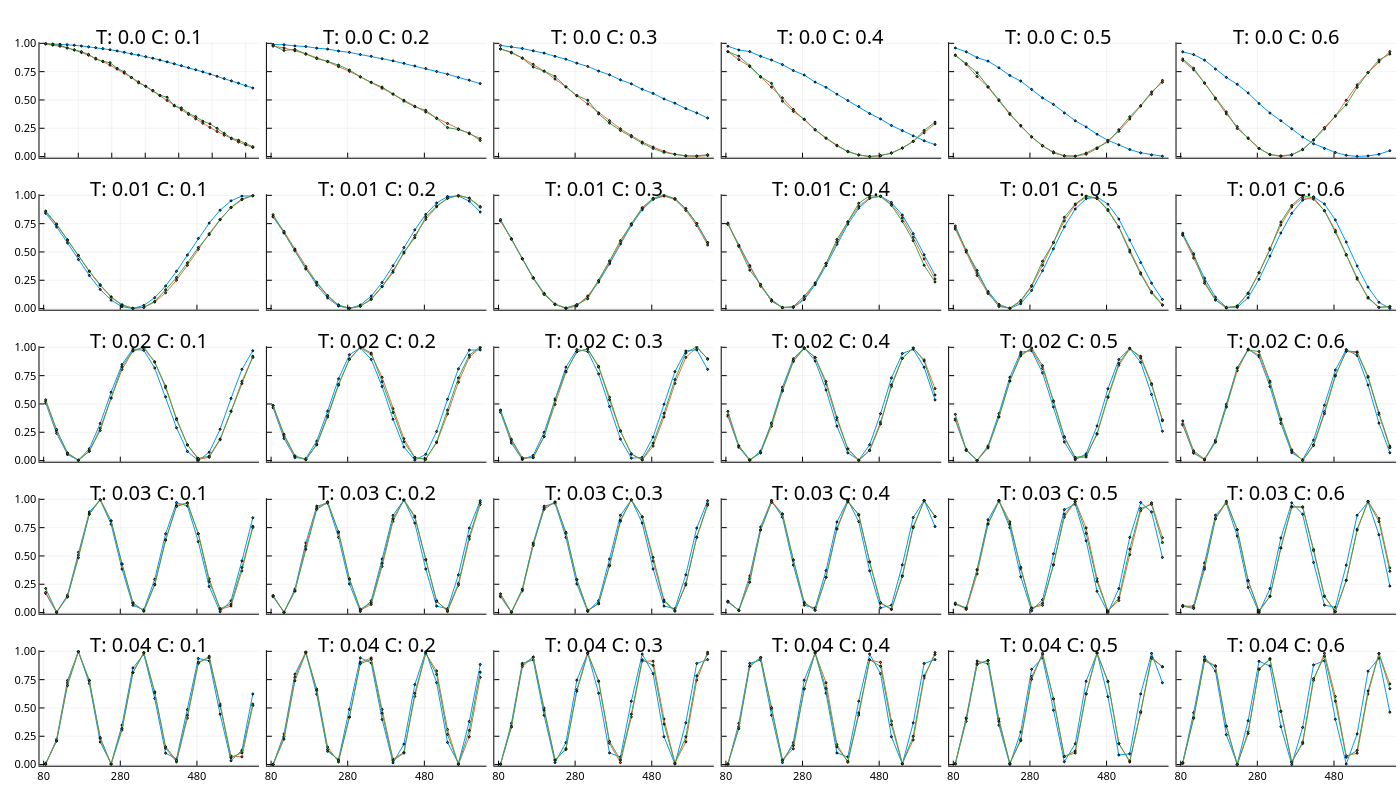}
    \caption{Plots of the survival probability of the initial state $\ket{00}$ as a function of time for different amplitudes for the pulses. }
    \label{fig:allplots00}
\end{figure}
\begin{figure}
    \centering
    \includegraphics[width=0.99\linewidth]{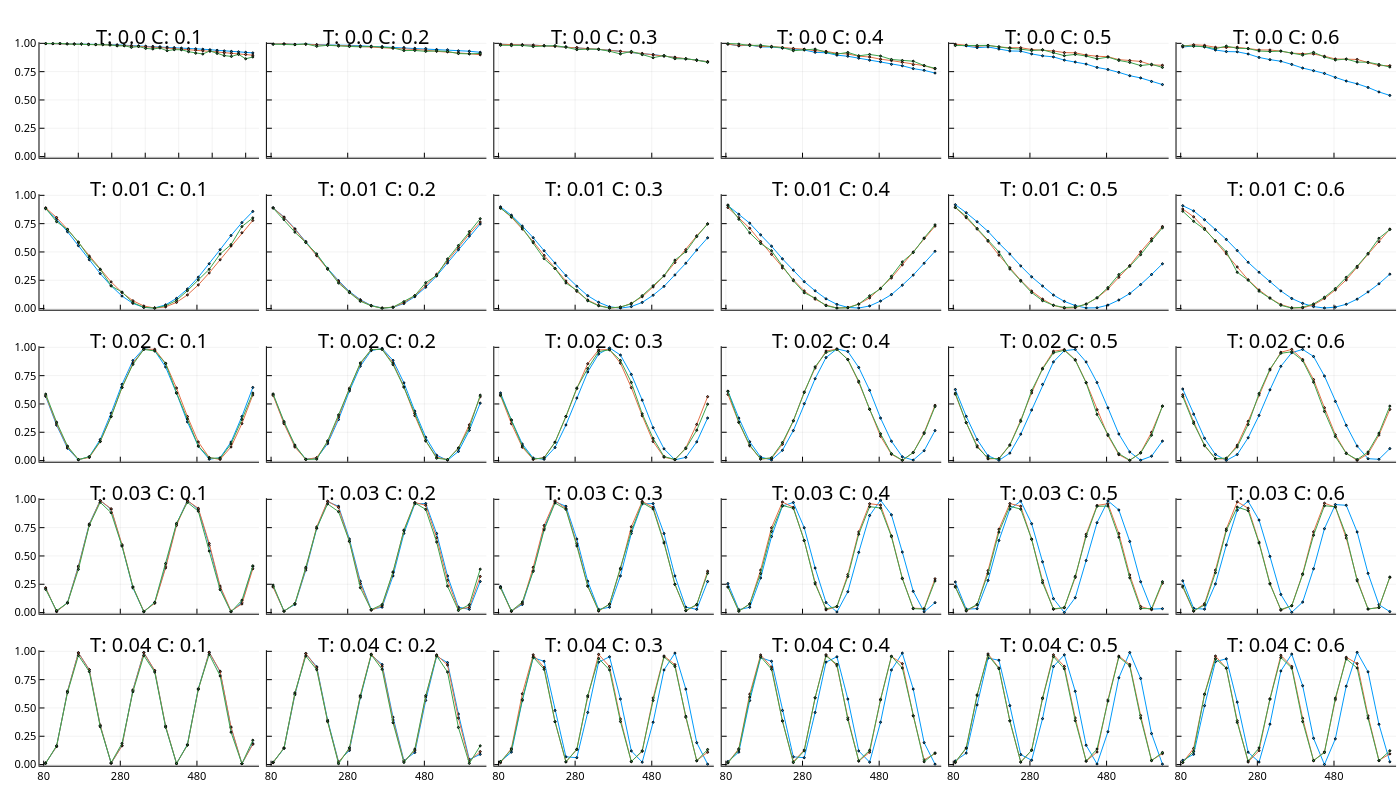}
    \caption{Plots of the survival probability of the initial state $\ket{10}$ as a function of time for different amplitudes for the pulses. }
    \label{fig:allplots10}
\end{figure}
\begin{figure}
    \centering
    \includegraphics[width=0.99\linewidth]{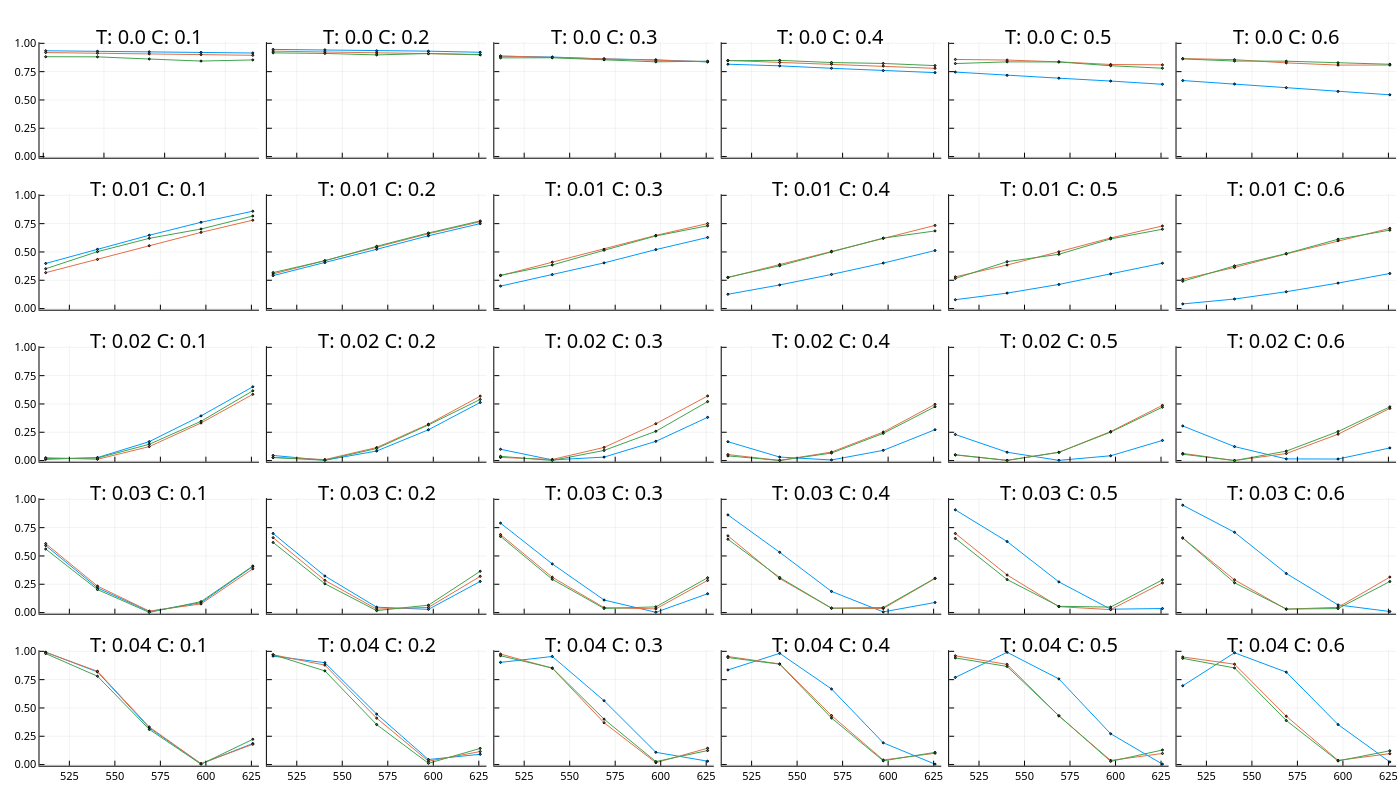}
    \caption{Plots of the survival probability of the initial state $\ket{11}$ as a function of time for different amplitudes for the pulses. }
\end{figure}
\begin{figure}
    \centering
    \includegraphics[width=0.99\linewidth]{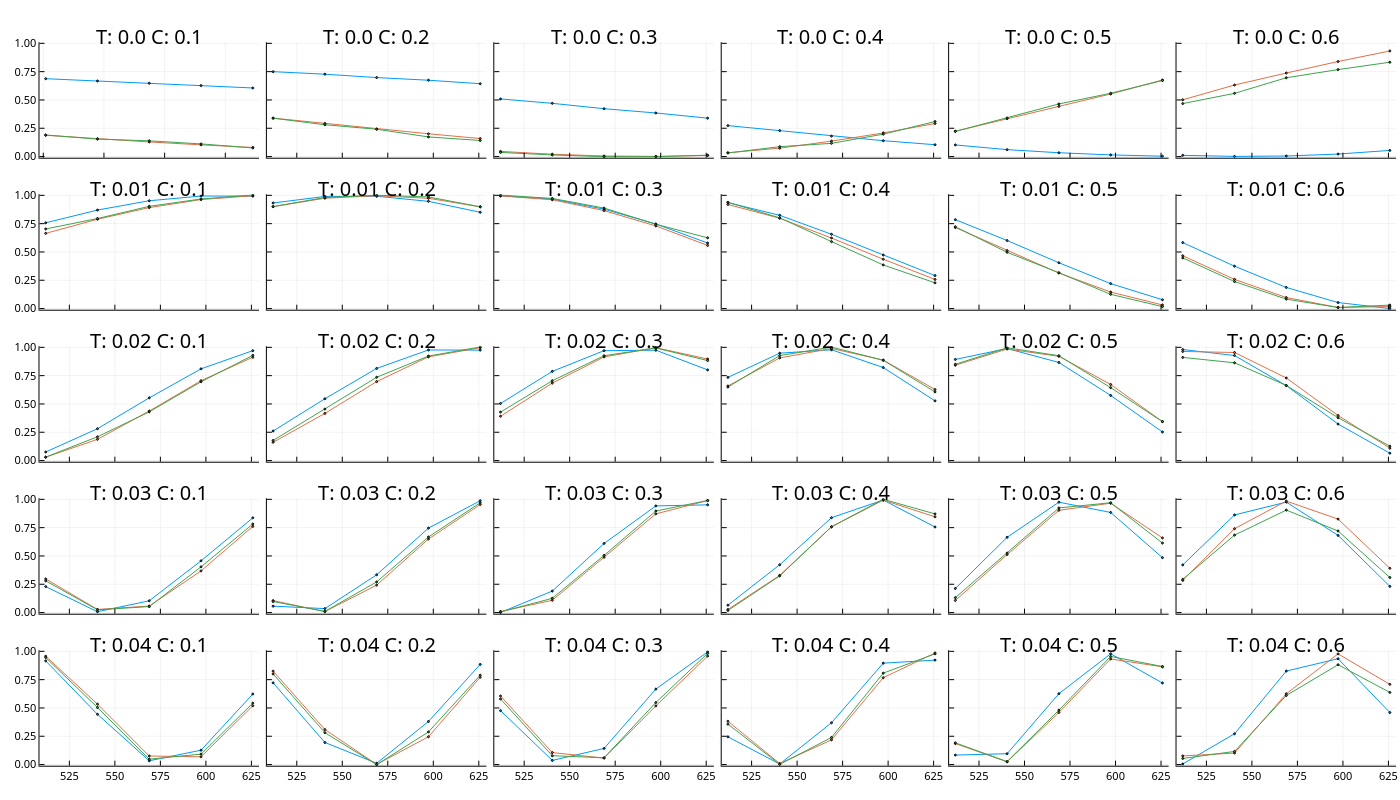}
    \caption{Plots of the survival probability of the initial state $\ket{01}$ as a function of time for different amplitudes for the pulses. }
\end{figure}

%% file: main.bbl
\providecommand{\newblock}{}
\begin{thebibliography}{10}
\expandafter\ifx\csname url\endcsname\relax
  \def\url#1{{\tt #1}}\fi
\expandafter\ifx\csname urlprefix\endcsname\relax\def\urlprefix{URL }\fi
\providecommand{\eprint}[2][]{\url{#2}}

\bibitem{Feynman1982}
Feynman R~P 1982 {\em International Journal of Theoretical Physics\/} {\bf 21} 467--488 ISSN 1572-9575

\bibitem{NielsenChuang}
Nielsen M~A and Chuang I~L 2010 {\em Quantum Computation and Quantum Information: 10th Anniversary Edition\/} (Cambridge University Press)

\bibitem{dawson2005solovaykitaevalgorithm}
Dawson C~M and Nielsen M~A 2005 The solovay-kitaev algorithm (\textit{Preprint} \eprint{quant-ph/0505030})

\bibitem{Quantum_Computers_nature}
Ladd T~D, Jelezko F, Laflamme R, Nakamura Y, Monroe C and O'Brien J~L 2010 {\em Nature\/} {\bf 464} 45--53 ISSN 1476-4687

\bibitem{josephsonjn_reproducibility}
Osman A, Simon J, Bengtsson A, Kosen S, Krantz P, P~Lozano D, Scigliuzzo M, Delsing P, Bylander J and Fadavi~Roudsari A 2021 {\em Applied Physics Letters\/} {\bf 118} 064002 ISSN 0003-6951

\bibitem{nv_quantum_computer}
Pezzagna S and Meijer J 2021 {\em Applied Physics Reviews\/} {\bf 8} 011308 ISSN 1931-9401

\bibitem{oneshottoffoli}
Zahedinejad E, Ghosh J and Sanders B~C 2015 {\em Phys. Rev. Lett.\/} {\bf 114}(20) 200502

\bibitem{IBMquantum}
Https://quantum.ibm.com/, 2021

\bibitem{qiskit_pulse}
Alexander T, Kanazawa N, Egger D~J, Capelluto L, Wood C~J, Javadi-Abhari A and C~McKay D 2020 {\em Quantum Science and Technology\/} {\bf 5} 044006 ISSN 2058-9565

\bibitem{rigetti}
Https://www.rigetti.com/

\bibitem{OQC}
Https://oqc.tech/

\bibitem{IQM}
Https://meetiqm.com/

\bibitem{QuantumCircuitsInc}
Https://quantumcircuits.com/

\bibitem{rl_control}
Bukov M, Day A~G~R, Sels D, Weinberg P, Polkovnikov A and Mehta P 2018 {\em Phys. Rev. X\/} {\bf 8}(3) 031086

\bibitem{rl_entangler}
Nam~Nguyen H, Motzoi F, Metcalf M, Birgitta~Whaley K, Bukov M and Schmitt M 2024 {\em Machine Learning: Science and Technology\/} {\bf 5} 025066 ISSN 2632-2153

\bibitem{comp_study}
Riaz B, Shuang C and Qamar S 2019 {\em Quantum Information Processing\/} {\bf 18} 100 ISSN 1573-1332

\bibitem{effective_cr}
Magesan E and Gambetta J~M 2020 {\em Phys. Rev. A\/} {\bf 101}(5) 052308

\bibitem{transmon}
Koch J, Yu T~M, Gambetta J, Houck A~A, Schuster D~I, Majer J, Blais A, Devoret M~H, Girvin S~M and Schoelkopf R~J 2007 {\em Phys. Rev. A\/} {\bf 76}(4) 042319

\bibitem{qiskit_dynamics}
Puzzuoli D, Lin S~F, Malekakhlagh M, Pritchett E, Rosand B and Wood C~J 2023 {\em Journal of Computational Physics\/} {\bf 489} 112262 ISSN 0021-9991

\bibitem{crossresonance}
Rigetti C and Devoret M 2010 {\em Phys. Rev. B\/} {\bf 81}(13) 134507

\bibitem{microwave_coupling}
Paraoanu G~S 2006 {\em Phys. Rev. B\/} {\bf 74}(14) 140504

\bibitem{tuning_up_cr}
Sheldon S, Magesan E, Chow J~M and Gambetta J~M 2016 {\em Phys. Rev. A\/} {\bf 93}(6) 060302

\bibitem{optimalcontrol}
Li X 2023 {\em Scientific Reports\/} {\bf 13} 14734 ISSN 2045-2322

\bibitem{torchqc}
Koutromanos D, Stefanatos D and Paspalakis E 2025 {\em Computer Physics Communications\/} {\bf 310} 109505 ISSN 0010-4655

\bibitem{qutip}
Lambert N, Giguère E, Menczel P, Li B, Hopf P, Suárez G, Gali M, Lishman J, Gadhvi R, Agarwal R, Galicia A, Shammah N, Nation P~D, Johansson J~R, Ahmed S, Cross S, Pitchford A and Nori F 2024 {QuTiP} 5: The quantum toolbox in {Python} (\textit{Preprint} \eprint{2412.04705})

\bibitem{grape}
Khaneja N, Reiss T, Kehlet C, Schulte-Herbrüggen T and Glaser S~J 2005 {\em Journal of Magnetic Resonance\/} {\bf 172} 296--305 ISSN 1090-7807

\bibitem{krotov}
Goerz M, Basilewitsch D, Gago-Encinas F, Krauss M~G, Horn K~P, Reich D~M and Koch C 2019 {\em SciPost physics\/} {\bf 7} 080

\bibitem{crab}
Caneva T, Calarco T and Montangero S 2011 {\em Phys. Rev. A\/} {\bf 84}(2) 022326

\bibitem{parametric_H}
Luchi P, Turro F, Quaglioni S, Wu X, Amitrano V, Wendt K, DuBois J~L and Pederiva F 2023 {\em The European Physical Journal A\/} {\bf 59} 196 ISSN 1434-601X

\bibitem{reinforcement_learning_CR}
Nam~Nguyen H, Motzoi F, Metcalf M, Birgitta~Whaley K, Bukov M and Schmitt M 2024 {\em Machine Learning: Science and Technology\/} {\bf 5} 025066

\bibitem{squeeze}
Robertson L~H~A 2023 squeeze: Accelerated quantum pulse schedules (\textit{Preprint} \eprint{2311.08742})

\bibitem{rackauckas2020universal}
Rackauckas C, Ma Y, Martensen J, Warner C, Zubov K, Supekar R, Skinner D and Ramadhan A 2020 {\em arXiv preprint arXiv:2001.04385\/}

\bibitem{SciMLBook}
Rackauckas C 2022 {\em Parallel Computing and Scientific Machine Learning (SciML): Methods and Applications\/} chap~11

\bibitem{NODE}
Chen T~Q, Rubanova Y, Bettencourt J and Duvenaud D 2018 {\em CoRR\/} {\bf abs/1806.07366} (\textit{Preprint} \eprint{1806.07366})

\bibitem{DifferentialEquations.jl}
Rackauckas C and Nie Q 2017 {\em The Journal of Open Research Software\/} {\bf 5} exported from https://app.dimensions.ai on 2019/05/05

\bibitem{sense_alg}
Ma Y, Dixit V, Innes M~J, Guo X and Rackauckas C 2021 A comparison of automatic differentiation and continuous sensitivity analysis for derivatives of differential equation solutions {\em 2021 IEEE High Performance Extreme Computing Conference (HPEC)\/} pp 1--9

\bibitem{fluxjl}
Innes M, Saba E, Fischer K, Gandhi D, Rudilosso M~C, Joy N~M, Karmali T, Pal A and Shah V 2018 {\em CoRR\/} {\bf abs/1811.01457} (\textit{Preprint} \eprint{1811.01457})

\bibitem{nesterov}
Nesterov Y 1983 {\em In {\em Soviet Mathematics Doklady}, volume~27, pages 372--376, 1983\/}

\bibitem{enzymejl}
Moses W and Churavy V 2020 Instead of rewriting foreign code for machine learning, automatically synthesize fast gradients {\em Advances in Neural Information Processing Systems\/} vol~33 ed Larochelle H, Ranzato M, Hadsell R, Balcan M~F and Lin H (Curran Associates, Inc.) pp 12472--12485

\end{thebibliography}
